\documentclass[12pt]{article}
\def\thefootnote{\fnsymbol{footnote}}
\newcommand{\vev}{{\it vev}}

\def\tF{{\tilde F}}
\def\bW{{\overline{\cal{W}}}}

\def\r{\right|}
\def\l{\left.}
\def\[{\left [}
\def\]{\right ]}
\def\({\left (}
\def\){\right )}
\def\|{\left |}

\def\rbr{\right\}}
\def\pp{\partial}

\def\ux{$U(1)_X$}
\def\ua{$U(1)_a$}

\def\t{\bar{t}}
\def\T{\bar{T}}

\def\im{{\rm Im}}
\def\re{{\rm Re}}
\newcommand{\beq}{\begin{equation}}
\newcommand{\eeq}{\end{equation}}
\newcommand{\bea}{\begin{eqnarray}}
\newcommand{\eea}{\end{eqnarray}}
\def\bF{\bar{F}}
\def\G{{\cal G}}

\def\L{{\cal L}}
\newcommand{\superint}{\int \diff^{4}\theta \, }
\newcommand{\superintF}{\int \diff^{4}\theta{E\over R} }

\newcommand{\diff}{\mbox{d}}

\def\bD{{\bar{\cal D}}}

\newcommand{\WaWa}{{\cal{W}}^{\alpha}{\cal{W}}_{\alpha}}

\newcommand{\DaDa}{{\cal D}^2}
\newcommand{\DbDb}{{\bar{\cal D}}^2}

\newcommand{\Wa}{{\cal W}^{\alpha}}

\newcommand{\Wc}{{\cal W}_{\alpha}}

\def\pp{\partial}

\def\Del{\Delta}

\def\del{\delta}

\newcommand{\myref}[1]{(\ref{#1})}
\newcommand{\beqa}{\begin{eqnarray}}
\newcommand{\eeqa}{\end{eqnarray}}
\newcommand{\nnn}{ \nonumber \\ }

\newcommand{\hc}{ + {\rm h.c.}}
\newcommand{\mhc}{ - {\rm h.c.}}

\newcommand{\chiproj}{(\bD^2 - 8R)}

\newcommand{\Zbf}{{{\bf Z}}}

\newcommand{\uone}{$U(1)$}

\newcommand{\myvev}[1]{{\langle #1 \rangle}}

\newcommand{\half}{{1 \over 2}}

\newcommand{\ddd}{\nnn && \quad}
\newcommand{\mmm}{\nnn & &}

 \newcommand{\F}{{\cal F}}

\begin{document}

\begin{titlepage}

\hfill   LBNL-56983

\hfill   UCB-PTH-05/04

\hfill   hep-th/0502100

\hfill \today

\begin{center}

\vspace{18pt}
{\bf The Axion Mass in Modular Invariant Supergravity}\footnote{This
work was supported in part by the
Director, Office of Science, Office of High Energy and Nuclear
Physics, Division of High Energy Physics of the U.S. Department of
Energy under Contract DE-AC03-76SF00098, in part by the National
Science Foundation under grant PHY-0098840.}

\vspace{18pt}

Daniel Butter\footnote{E-Mail: {\tt dbutter@socrates.Berkeley.EDU}}
\vskip .01in
{\em Department of Physics, University of California, Berkeley,
CA 94720 USA}
\vskip .03in
{\em and}
\vskip .03in
Mary K. Gaillard,\footnote{E-Mail: {\tt MKGaillard@lbl.gov}}
\vskip .01in
{\em Department of Physics, University of California 
and \\ Theoretical Physics Group, Bldg. 50A5104,
Lawrence Berkeley National Laboratory \\ Berkeley,
CA 94720 USA}

\vspace{18pt}

\end{center}

\begin{abstract} When supersymmetry is broken by condensates
with a single condensing gauge group, there is a nonanomalous
R-symmetry that prevents the universal axion from acquiring a mass.
It has been argued that, in the context of supergravity, higher
dimension operators will break this symmetry and may generate an axion
mass too large to allow the identification of the universal axion with
the QCD axion.  We show that such contributions to the axion mass are
highly suppressed in a class of models where the effective Lagrangian
for gaugino and matter condensation respects modular invariance
(T-duality).

\end{abstract}

\end{titlepage}

\newpage
\renewcommand{\thepage}{\roman{page}}
\setcounter{page}{2}
\mbox{ }

\vskip 1in

\begin{center}
{\bf Disclaimer}
\end{center}

\vskip .2in

\begin{scriptsize}
\begin{quotation}
This document was prepared as an account of work sponsored by the United
States Government. Neither the United States Government nor any agency
thereof, nor The Regents of the University of California, nor any of their
employees, makes any warranty, express or implied, or assumes any legal
liability or responsibility for the accuracy, completeness, or usefulness
of any information, apparatus, product, or process disclosed, or represents
that its use would not infringe privately owned rights. Reference herein
to any specific commercial products process, or service by its trade name,
trademark, manufacturer, or otherwise, does not necessarily constitute or
imply its endorsement, recommendation, or favoring by the United States
Government or any agency thereof, or The Regents of the University of
California. The views and opinions of authors expressed herein do not
necessarily state or reflect those of the United States Government or any
agency thereof of The Regents of the University of California and shall
not be used for advertising or product endorsement purposes.
\end{quotation}
\end{scriptsize}

\vskip 2in

\begin{center}
\begin{small}
{\it Lawrence Berkeley Laboratory is an equal opportunity employer.}
\end{small}
\end{center}

\newpage
\renewcommand{\thepage}{\arabic{page}}
\setcounter{page}{1}
\def\thefootnote{\arabic{footnote}}
\setcounter{footnote}{0}

Banks and Dine~\cite{bd} pointed out ten years ago that in a
supersymmetric Yang Mills theory with a dilaton chiral superfield that
couples universally to Yang-Mills fields:
\beq \L_{YM} = {1\over8}\sum_a\int d^2\theta
S(\WaWa)_a\hc,\label{bd}\eeq
there is a residual R-symmetry in the effective theory for the
condensates of a strongly coupled gauge sector, provided that there is
a single condensation scale governed by a single $\beta$-function,
there is no explicit R-symmetry breaking by fermion mass terms in the
strongly coupled sector, the dilaton $S$ has no superpotential
couplings, and the K\"ahler potential is independent of $\im S$. The
latter two requirements are met in effective supergravity obtained
from the weakly coupled heterotic string, and explicit realizations of
this scenario can be found in the BGW model~\cite{bgw} and
generalizations~\cite{ggm} thereof to include an anomalous \ux.

The R-symmetry transformations on the gauginos $\lambda_a$ and chiral
fermions $\chi^A$:
\beq \lambda_a\to e^{{i\over2}\alpha}\lambda_a, \qquad \chi^A\to
e^{-{i\over2}\alpha}\chi^A,\label{rferm}\eeq
leave the classical Lagrangian \myref{bd} invariant, but are anomalous
at the quantum level:
\beq \Del\L_{YM} = {i\alpha\over8}\sum_a b'_a\int d^2\theta(\WaWa)_a\hc,\qquad
b'_a = {1\over8\pi^2}\(C_a - \sum_A C^A_a\),\label{delbd}\eeq
where $C_a$ and $C^A_a$ are quadratic Casimir operators in the adjoint
and matter representations, respectively. In the case that there is
a single simple gauge group $\G_c$ the symmetry can be restored by
an axion shift:
\beq a = \l\im S\r \to a - ib'_c\alpha.\eeq
If this gauge group becomes strongly coupled at a scale 
\beq \Lambda_c \sim e^{-1/3b_c g^2_0}\Lambda_0, \qquad b_a =
{1\over8\pi^2}\(C_a - {1\over3}\sum_A C^A_a\),\label{lambda}\eeq
the effective theory~\cite{vyt} below that scale will have the same
anomaly structure as the underlying theory.   A potential is generated
for the dilaton $d = \l\re S\r$, but not for the axion.  If the gauge
group is not simple: $\G = \prod_a\G_a$, the R-symmetry is anomalous,
but no mass is generated for the axion as long as there is a single
condensate.  In the two condensate case with $\beta$-functions
$b_2\ll b_c$ for the models of~\cite{bgw,ggm} the axion acquires
a small mass:
\beq m_a \sim (\Lambda_2/\Lambda_c)^{3\over2}m_{3\over2}.\label{ma}\eeq
In the context of the weakly coupled heterotic string a viable
scenario for supersymmetry breaking occurs if a hidden sector gauge
group condenses with~\cite{gnnb} $b_c\approx .03$, $\Lambda_c\sim
10^{13}$GeV, $m_{3\over2}\sim$ TeV.  Then if there is no additional
condensing gauge group other than QCD, the universal axion is a
candidate Peccei-Quinn axion with mass
\beq m_a \sim 10^{-9}{\rm eV},\label{ma2}\eeq
as suggested\footnote{The result \myref{ma} cannot be directly applied
to the QCD axion, since QCD condensation occurs far below the scale
of supersymmetry breaking and heavy modes need to be correctly
integrated out.} by \myref{ma}.  Note that this mass is decoupled from
the axion coupling constant, which in these models\footnote{If the
classical dilaton K\"ahler potential is used, the axion couping
constant is approximately~\cite{fpt} $10^{16}$ Gev. The BGW model
invokes string nonperturbative corrections to stabilize the dilaton.
These have the effect of dramatically enhancing the dilaton mass and
moderately enhancing the axion coupling constant: $F_a \approx
(\sqrt{6}/b_c)\times10^{16}$ Gev $\approx 6\times 10^{17}$ Gev.  In
the third paper of~\cite{bgw} it was incorrectly stated that the axion
coupling constant was suppressed by these effects.} is of the order of
the reduced Planck mass $m_P = 1/\sqrt{8\pi G_N}$. As a result,
analyses~\cite{fpt} of the viability of such an axion must be
revisited.

However Banks and Dine also pointed out~\cite{bd} that in the context
of supergravity one would expect higher order terms to be generated;
terms of the form 
\beq \L' = {1\over8}\sum_n\lambda_n\int
d^2\theta(\WaWa)^n\hc,\label{wwn}\eeq
do not respect R-symmetry for $n>1$.  Supergravity is more
restrictive than global supersymmetry; in the language of K\"ahler
\uone\, supergravity~\cite{bgg}, superpotential terms $W_i$ must have K\"ahler
\uone\, weight 2, where chiral fields $\Phi^A$ have weight 0 and the
Yang-Mills superfield strength $\Wc$ has weight 1.  Thus the following
terms with at least one factor $\WaWa$ are allowed
\beq \L_{SP} = {1\over2}\superintF\WaWa\F(e^{-K/2}\WaWa,Z^A)
\hc,\label{lsp}\eeq
where $E$ is the superdeterminant of the supervielbein, $R$ is an
element of the superspace curvature tensor, and $Z^A$ is any chiral
superfield.  Effective supergravity from the weakly coupled heterotic
string is perturbatively invariant~\cite{mod} under T-duality
transformations that, in the class of models
studied in~\cite{bgw,ggm}, take the form
\bea T^I &\to& {a^I T^I - i b^I \over ic^I T^I + d^I}, \qquad \Phi^A
\to e^{i\del_A -\sum_I q_I^A F^I} \Phi^A,\nnn \lambda_L&\to&
e^{-{i\over2}\im F}\lambda_L,\qquad F^I = \ln \( i c^I
T^I + d^I \),\nnn a^I d^I - b^I c^I &=& 1, \qquad a^I,b^I,c^I,d^I \in
\Zbf \qquad \forall \quad I=1,2,3,
\label{mdtr}
\eea
and under which the K\"ahler potential and superpotential 
transform as 
\beq K\to K + F + \bF, \qquad W\to e^{- F}W, \qquad F = \sum_I
F^I.\eeq
Here the $T^I$ are gauge neutral K\"ahler moduli, and the moduli
independent~\cite{lust,fer} phases $\del^A$ depend on the parameters
$a^I,b^I,c^I,d^I$ of the transformation and on the modular weights
$q^A_I$.  Modular invariance then further restricts the superpotential
couplings as follows
\beq \F = \F(\eta^2e^{-K/2}\WaWa,\eta^A\Phi^A), \qquad
 \eta = \prod_I\eta_I,\qquad \eta^A = \prod_I\eta_I^{2q^A_I},
\qquad \eta_I = \eta(i T^I),\label{hsp}\eeq
with the Dedekind functions transforming under \myref{mdtr} as
\beq \eta(i T^I)\to e^{i\del_I}e^{\half F(T^I)}\eta(i T^I),\qquad F(T^I) =
F^I,\qquad \del_I = \del_I(a^I,b^I,c^I,d^I).\label{phase}\eeq
Consider first terms with no $\Phi^A$-dependence; since~\cite{SW} for
a general transformation \myref{mdtr} $\del_I = n_I\pi/12$, the
only invariant superpotential is of the form:
\beq \L_{H W} = {1\over2}\superint{E\over R}\WaWa\F(\eta^2e^{-K/2}\WaWa)
\hc,\qquad \F(X) = \sum_{n=1}\lambda_n X^{12n}.\label{lhw}\eeq
If the $[SL(2,{\bf Z})]^3$ symmetry implied by \myref{mdtr} were instead
restricted, say to just $SL(2,{\bf Z})$, with $a_I,b_I,c_I,d_I,$
independent of $I$, then the phase of $\eta$ in \myref{phase} is
$3\del_I = n\pi/4$, and lower dimension operators would be allowed:
$\F(X) = \sum_{n=1}\lambda_n X^{4n}$, which according to the estimate
of~\cite{bd} is of sufficiently high dimension to avoid an unacceptably
large mass for the QCD axion.  We can explicitly calculate this mass
in the BGW model.

To construct an effective theory below the scale of gaugino
condensation, one has to introduce~\cite{bg89}
a chiral superfield of chiral weight 2:
\beq \WaWa\sim U\sim e^{K/2}H^3,\eeq
where $H$ is an ordinary chiral superfield of zero chiral weight and
dimension one.  The most straightforward way to implement this
requirement is to put the dilaton in a vector supermultiplet and
impose~\cite{bdqq,bgt}
\bea U&=&-(\DbDb-8R)V, \qquad
\bar{U}=-(\DaDa-8R^{\dagger})V. \label{modlin1} \eea
This parallels the modified linearity condition for the underlying
field theory in the dual (and in fact string derived)
formulation with the dilaton as the lowest component of a linear
supermultiplet whose components include a two-form potential $b_{m n}$
dual to the axion.  This formalism has the advantages that
the Bianchi identity
\beq (\DaDa-24R^{\dagger})U - (\DbDb-24R)\bar{U} ={\rm{total\;
derivative.}}\label{uconst} \eeq
is automatically satisfied, and that when the Green-Schwarz term
needed to cancel the field theoretic modular anomaly is included,
there is no mixing of the dilaton with the K\"ahler moduli $T^I$.
In this formulation the axion shift is traded for a two-form
gauge symmetry: $b_{m n}\to b_{m n} + \nabla_{[m}\Lambda_{n]}$.
Since only the gauge invariant 3-form $h_{m n r} =
\nabla_{[m}b_{n r]}$ appears in the Lagrangian, the role of this
symmetry is less apparent.  We will explicitly calculate the
modification of the scalar potential in the presence of a term of the 
form \myref{lhw} with $\WaWa\to U$ and $\F = \lambda(\eta^2e^{-K/2}U)^p
= \lambda X^p$.

The BGW Lagrangian~\cite{bgw,ggm} is given by

\beq \L = \superint E\[-3 + V\(2s(V) + V_{GS}\)\] + \L_{VYT} + \L_{th},\eeq
where $s(\myvev{\ell}) = g_s^{-2}$, with $\ell = \l V\r$ and $g_s$ the
string scale gauge coupling constant, $V_{GS}$ is the four dimensional
analogue of the Green-Schwarz counterterm needed to cancel
modular~\cite{gsm} and \uone\,~\cite{gsu} anomalies, $\L_{VYT}$ is the
``quantum'' part of the condensate Lagrangian, constructed by standard
anomaly matching~\cite{vyt} to the quantum-induced
correction~\cite{quant} in the underlying theory, and $\L_{th}$ is the
string-loop correction~\cite{thresh} to the Yang-Mills coupling.  Upon
solving the equations of motion for the auxiliary fields and the
(static\footnote{The dynamical condensate case was studied
ref.~\cite{yy} for an $E_8$ gauge condensate without
matter. It was found that both the condensate magnitude $\rho$ and its
phase $\omega$ have masses larger than the condensation scale. After
integrating out these fields, one recovers the theory with a static
$E_8$ condensate studied in the first paper in~\cite{bgw}.})
condensates, the relevant part of the scalar Lagrangian takes the form
(up to a total derivative)
\bea e^{-1}\L &=& -\half r - (1 + b\ell)\sum_I{\pp_m\t^I\pp^m
t^I\over(2\re t^I)^2} - {k'(\ell)\over4\ell}\pp^m\ell\pp_m\ell - V +
\L_a,\nnn V &=& {|u|^2\over16}\(\ell k'(\ell)\left|\ell^{-1} + b_c +
4X\F'(X)\r^2 - 3\left|b_c + 4X\F'(X)\r^2\right.\mmm \l + (1 +
b\ell)\left|b - b_c - 4X\F'(X)\r^2\sum_I\left|1 + 4\re
t^I\zeta_I)\r^2\rbr,\nnn \L_a &=& {k'\over4\ell}B^m B_m + i
B_m\sum_I{b\over4\re t^I}\[\(1 + 4\re t^I\zeta_I\) \pp^m t^I\mhc\]\ddd
- \[b_c\omega - 2i\(\F + X\F'\mhc\)\]\nabla^m B_m, \label{lag}\eea
where
\bea u &=& \l U\r = |u|e^{\omega_0}, \qquad \omega = \omega_0 -
i\sum_I \ln(\eta_I/\bar\eta_I), \qquad X = e^{-K/2}u\eta^2 =
x(\ell,y_I,\omega) e^{i\omega},  \nnn y_I &=& |\eta_I|^4(t^I+\t^I),
\qquad \pp_I y_I = |\eta_I|^4\(1 + 4\re t^I\zeta_I\), \qquad \zeta_I =
{\pp\ln\eta_I\over\pp t^I}.\eea
The expression for the real function $x$ is determined by the equation
of motion for the real part $(F_U + \bF_U)/2$ of the auxiliary field of the
superfield $U$:
\beq |u|^2 = e^{k(\ell)}x^2/\prod_I y_I = \rho^2(\ell,y_I)\exp\[-4
\(\F + X\F'\hc\)\],\eeq
where $\rho$ is the solution for $|u|$ found in~\cite{bgw} with $\F = 0$.
The one-form $B_m$ is dual to a three-form:
\beq B_m = \half\epsilon_{m n p q}\({1\over3!4}\Gamma^{n p q} + \pp^n
b^{p q}\),\label{Bm} \eeq
with $\Gamma$ and $b$ 3-form and 2-form potentials, respectively.
If $\nabla^m B_m = 0$, $\Gamma = 0$, and $b_{m n}$ is dual to a massless
scalar. This is the case when $\F=0$, in which case the equation of motion
for $\omega$ is $b_c\nabla^m B_m = - \pp V/\pp\omega = 0.$  In the
presence of the terms $\F = \lambda X^p$ the potential is no longer independent
of $\omega$ and its equation of motion gives $\nabla^m B_m =
\epsilon_{m n p q}\pp^m\Gamma^{n p q}/3!8\ne 0.$ In this case the 2-form
$b$ can be removed by a gauge transformation $\Gamma^{n p q}\to
\Gamma^{n p q} - 3!4\pp^n b^{p q}$, and the equation of motion for
$\Gamma$ is just $\pp L/\pp B_m = 0.$  Setting the moduli $t^I$ at
self-dual points $t^I = t_{sd}$, which minimize the potential and satisfy
$\pp_I y_I = 0$, and retaining only leading order terms in the correction
$\F$, the relevant equations of motion are:
\bea {\del\L\over\del\omega} &\approx& - b_c\nabla^m B_m - {\pp
V\over\pp\omega} = 0,\nnn {\del\L\over\del B_m} &\approx&
{k'\over2\ell}B^m + b_c\pp^m\omega, \qquad 2b_c^2\nabla^m\({\ell\over
k'}\pp_m\omega\) \approx {\pp V\over\pp\omega},\label{eoms}\eea
which are equivalent to the equation of motion for the scalar $\omega$
with the Lagrangian $\L_a$ replaced by (neglecting $t^I$)
\beq \L_a(\omega) = - {b_c^2\ell\over k'}\pp^m\omega\pp_m\omega.\eeq
The normalized mass of the axion is given by 
\beq m^2_a = {k'\over2b_c^2\ell}{\pp^2V\over\pp\omega^2} \approx
{p^3|u|^2k'\lambda|X|^p\over4b_c^2\ell}\[3b_c - (1 + b_c\ell)
{k'}\]\approx 36p^3\lambda|X|^p b_c^{-1}
m^2_{3\over2}\label{amass}, \eeq
where 
\beq m_{3\over2} \approx {b_c|u|\over4} \eeq
is the gravitino mass, and we used the fact that (near) vanishing of
the cosmological constant requires~\cite{bgw}
$\myvev{\ell^{-1}k'(\ell)} \approx 3b_c^2\ll1$.  If, say, $b_c \approx
.03$, $m_{3\over2}\approx$ TeV, $|X|\approx |u|\approx 10^{-13}$ in
reduced Planck units,\footnote{More precisely, $|X| =
e^{-k/2}u\prod_I(y_I)^{\half}$ with $\myvev{y_I}\approx .7$ at
the self-dual points, and we generally expect
the factor $\myvev{e^{-k/2}}$ to be smaller that its classical value
$\myvev{\ell^{-\half}} = \sqrt{2}/g_s \approx 2 .$} $\lambda\approx1$,
this gives $m_a \approx 10^{-12}$eV ($10^{-63}$eV) if $p = 4(12)$.

In order to obtain the axion couplings to unconfined gauge superfields
$\Wa$ we have to include them in the modified linearity condition
\myref{modlin1} which then reads
\bea U + \WaWa &=&-(\DbDb-8R)V, \qquad \bar{U} +
\bW_{\dot\alpha}\bW^{\dot\alpha}
=-(\DaDa-8R^{\dagger})V. \label{modlin2} \eea
Then $\L_a$ in \myref{lag} is replaced by
\bea\L_a &=&
{k'\over4\ell}H^m H_m + i H_m\sum_I{b\over4\re t^I}\[\(1 + 4\re
t^I\zeta_I\) \pp^m t^I\mhc\]\ddd - \[b_c\omega - 2i\(\F +
X\F'\mhc\)\]\nabla^m B_m,\nnn H_m &=& B_m + \omega_m,\label{lag2}\eea
where $\omega_m$ is dual to the Yang-Mills Chern-Simons 3-form,
normalized such that
\beq \nabla^m\omega_m = {1\over4}F\cdot\tF,\qquad
\nabla^m H_m = \nabla^m B_m + {1\over4}F\cdot\tF \eeq
and $B_m$ is decomposed as in \myref{Bm}.  If $\F=0$, then $\pp
V/\pp\omega=0$ and the equation of motion for $\omega$ gives $\nabla^m
B_m = 0,$ $\Gamma = 0$. Setting the $t^I$ at self-dual points, the
equation of motion for the two-form $b_{m n}$ gives
\beq \epsilon_{m n p q}\nabla^p\({k'\over2\ell}H^q\) = 0, \qquad
{k'\over2\ell} H^p = \pp^p a, \qquad
\nabla^m H_m = {1\over4}F\cdot\tF = \nabla^m\({2\ell\over k'}\pp_m a\),\eeq
which is the equation of motion for a massless axion $a$ with
Lagrangian
\beq \L_a(a) = - {\ell\over k'}\pp^m a\pp_m a - {a\over4}F\cdot\tF
\label{Laa}.\eeq
With $\F\ne0$ and $\pp
V/\pp\omega\ne0$, the equation of motion for $\omega$ gives the
first line of \myref{eoms}, and the second line is replaced by
\beq {\del\L\over\del B_m} \approx {k'\over2\ell}H^m +
b_c\pp^m\omega, \qquad 2b_c\nabla^m\({\ell\over k'}\pp_m\omega\) \approx
{\pp V\over b_c\pp\omega} - {1\over4}F\cdot\tF.\label{eoms2}\eeq
Setting $a = - b_c\omega$, the equivalent axion Lagrangian is
$e\[\L_a(a) - V(a,\ell,t^I)\]$, with the axion mass given by
\myref{amass}.

In addition to the operators in \myref{hsp} chiral superfields with
zero chiral weight can be constructed using chiral projections of any
functions of chiral fields.  Operators of this type were
found~\cite{agnt} in (2,2) orbifold compactifications of the heterotic
string theory with six dynamical moduli.\footnote{The results
of~\cite{agnt} are presented in the superconformal formalism of
supergravity with conformal gauge fixing by a chiral compensator that
plays an analagous to the K\"ahler weight factor in \myref{lsp}.}
In the class of models considered here we can construct zero-weight
chiral superfields of the form
\beq\F =
e^{-(p+n)K/2}(e^{-K/2}\WaWa)^p\eta^{2(p+n)}\prod_{i=1}^n\chiproj
f_i(y_I)\label{chiproj}\eeq
that are modular invariant provided $(p+n)\sum_I\del = m\pi.$ Since
$\myvev{F^I}=0$, the corresponding terms in the potential at the
condensation scale are proportional to\footnote{The coefficients of
the nonpropagating condensate superfield auxiliary fields vanish by
their equations of motion.} $|u|^p(m_{3\over2})^{n+1}$, so for fixed
$p+n$ one is trading factors of $|u|$ for factors of $m_{3\over2}\sim
10^{-2}|u|$, and these contributions to the axion mass will be smaller
than those in \myref{amass}.

We may also consider operators with matter fields that have
nonvanishing \vev's.  Since $e^{-K/2}\WaWa$ transforms like the
composite operators $U_1U_2U_3$ constructed from untwisted chiral
superfields, the rules for construction of a covariant superpotential
including this chiral superfield can be directly extracted from the
discussion in~\cite{mk} of modular invariant superpotential terms in
the class of $Z_3$ orbifolds considered here.  They take the form of
\myref{hsp} with
\beq \F_{p n q} = \Pi^q(e^{-K/2}\WaWa)^p\eta^{2(p +n)}\prod_{\alpha =
1}^n W_i, \qquad \Pi = Y^1Y^2Y^3, \qquad (p+n)\sum_I\del_I =
m\pi,\label{fpnq}\eeq
where $Y^I$ is a twisted sector oscillator superfield, and $W_i$
is any modular covariant [$W_i\to e^{-F}W_i$ under
\myref{mdtr}] zero-weight chiral superfield that is a candidate
superpotential term, subject to other constraints, such as gauge
invariance.  For example,the superpotential terms for matter
condensates could contribute to this expression.  However the
equations of motion for the auxiliary fields of these condensates give
$W_i\sim m_{3\over2}$ for these terms, so again they are less
important than the contribution in \myref{amass}.

Most $Z_3$ orbifold compactifications of the type considered here
have~\cite{joel} a \uone\, gauge group, denoted \ux, that is anomalous
at the quantum level of the effective field theory.  The anomaly is
canceled by a Green-Schwarz counterterm that amounts to a
Fayet-Illiopoulos D-term~\cite{gsu}.  A number $n$ of scalars $\phi^A$
acquire \vev's along an F- and D-flat direction such that $m\le n$
\ua\, gauge factors are broken at a scale $\Lambda_D$ that is close to
the Planck scale.  {\it A priori} there might be gauge and modular
invariant monomials of the form \myref{fpnq} with considerably larger
\vev's than those in \myref{lhw}, and no modular covariant, gauge
invariant superpotential term $W_i$, so that the direction $\phi^A\ne0$ is F-flat.
However if $m=n$, there is no gauge
invariant monomial $\prod_A(\phi^A)^{p_A}$.  Gauge invariance requires
\beq \sum_A p_A q^a_A = 0 \qquad \forall a,\label{gauge}\eeq
where $q^a_A$ is the \ua\, charge of $\phi^A$. If $m=n$ these are linearly
independent and form an $m\times m$ matrix with inverse $Q^A_a$; then
\myref{gauge} implies
\beq p_A = 0 \qquad \forall A.\label{gauge2}\eeq
Similarly, for the chiral projection of a monomial
$\prod_A(\phi^A)^{p_A + q_A}(\bar\phi^{\bar A})^{q_A}$ gauge
invariance still requires \myref{gauge} and \myref{gauge2}, so any
such monomial can be written in the form 
\beq f(T^J,\T^{\bar J}) \prod_A\[|\phi^A|^2\prod_I(T^I + \T^{\bar
I})^{-q^A_I}\]^{q_A}.\eeq
It is the modular invariant composite fields $|\phi^A|^2\prod_I(T^I +
\T^{\bar I})^{-q^A_I}$ that acquire~\cite{ggm} large \vev's; any
coefficients of them appearing in overall modular invariant operators
are subject to the same rules of construction as the operators in
\myref{chiproj}.  The same considerations hold if $N$ sets of fields
$\phi^A_i$ with identical \ua\, charges $(q^i_A)^a = q^a_A$, $i =
1,\ldots,N$ acquire \vev's.  This is the class of ``minimal'' models
studied in~\cite{ggm}; the dilaton potential in this class is
identical to that of the BGW model.

In the case $n>m$ one cannot rule out the above terms.  However in
this case, charge assignments that satisfy \myref{gauge} for $p_A>0$,
as in a holomorphic monomial, tend to destabilize~\cite{ggm} the
potential in a direction where the dilaton K\"ahler metric goes
negative and are therefore disfavored.  Moreover, in this case part of
the modular symmetry is realized nonlinearly on the \ua-charged
scalars after \ua-breaking. Monomials of the above type would generate
mixing of the axion with massless ``D-moduli'' that are Goldstone
particles~\cite{bdfs} associated with the degeneracy of the vacuum at
the \ua-breaking scale, requiring a more careful analysis.

In conclusion, modular invariant $Z_3$ models for gaugino condensation
with no \uone-breaking or with \uone-breaking by a minimal set of
scalar fields have highly suppressed contributions to the axion mass
from higher dimension operators.  Following~\cite{bgw} we have used
the linear multiplet formalism for the dilaton supermultiplet.
However, we expect~~\cite{bdqq,gn} that these results can be reproduced
in the chiral multiplet formalism.

\vskip 0.20in
\noindent {\bf \Large Acknowledgments}

\vspace{5pt}

\noindent 
We wish to thank David Lyth, Brent Nelson and Tom Taylor for helpful
discussions.  This work was supported in part by the Director, Office
of Science, Office of High Energy and Nuclear Physics, Division of
High Energy Physics of the U.S. Department of Energy under Contract
DE-AC03-76SF00098, in part by the National Science Foundation under
grant PHY-0098840.


\begin{thebibliography}{99}
%
\bibitem{bd} T. Banks and M. Dine, {Phys. Rev.} D {50}
(1994) 7454. 
%
\bibitem{bgw}
P. Bin\'etruy, M. K. Gaillard and Y.-Y. Wu,
Nucl. Phys. B 481 (1996) 109 and 493 (1997) and
Phys. Lett. B 412 (1997) 288.
%
\bibitem{ggm} M. K. Gaillard and J. Giedt, Nucl. Phys. B 636 (2002) 365
and Nucl. Phys. B 643 (2002) 201;
M. K. Gaillard, J. Giedt, and A. Mints, Nucl. Phys. B 700 (2004) 205.
%
\bibitem{vyt} G. Veneziano and S. Yankielowicz, {Phys. Lett. { B113}}
(1982) 231; T.R. Taylor, G. Veneziano and S. Yankielowicz,
Nucl. Phys. {B 218} (1983) 493; D. L\"ust and
T.R. Taylor. Phys. Lett. {B 253} (1991) 335; T.R. Taylor,
{Phys. Lett. {B 164}} (1985) 43.
%
C\bibitem{gnnb} M.K. Gaillard and B. Nelson, {Nucl. Phys.} B 571 (2000)
3; A. Birkedal-Hansen, B. D. Nelson, Phys. Rev. D 64 (2001) 015008.
%
\bibitem{fpt} P. Fox, A. Pierce and S. Thomas, hep-th/0409059 .
%
\bibitem{bgg}
P. Bin\'{e}truy, G. Girardi, R. Grimm,
Phys. Rep. 343 (2001) 255.
%
\bibitem{mod} A. Giveon, N. Malkin and E. Rabinovici, {
  Phys. Lett.}  {B220} (1989) 551; E. Alvarez and M. Osorio, {
  Phys. Rev.} {D40} (1989) 1150.
%
\bibitem{lust} S. Ferrara, D. L\"ust, A. Shapere and S. Theisen,
  Phys. Lett. B {225} (1989) 363.
%
\bibitem{fer} S. Ferrara, D. L\"ust and S. Theisen, Phys. Lett. B
  {233} (1989) 147.
%
\bibitem{SW} A. N. Schellekens and N. P. Warner, Nucl. Phys. {\bf B287},
317 (1984).
%
\bibitem{bg89} P. Bin\'{e}truy and M.K. Gaillard, 
{Phys. Lett. {B 232}} (1989) 82. 
%
\bibitem{bdqq} C.P. Burgess, J.-P. Derendinger, F. Quevedo and
M. Quir\'{o}s, {Phys. Lett. B 348 } (1995) 428. 
%
\bibitem{bgt} P. Bin\'{e}truy,
M.K. Gaillard and T.R. Taylor, {Nucl. Phys. B 455} (1995) 97.
%
\bibitem{gsm} G. L. Cardoso and B. A. Ovrut, Nucl. Phys. B 369 (1992)
351; J.-P. Derendinger, S. Ferrara, C. Kounnas and F. Zwirner,
Nucl. Phys. B 372 (1992) 145.
%
\bibitem{gsu} M. Dine, N. Seiberg, E. Witten, Nucl. Phys. B 289 (1987)
585; J. J. Atick, L. Dixon, A. Sen, Nucl. Phys. B 292 (1987) 109;
M. Dine, I. Ichinose, N. Seiberg, Nucl. Phys. B 293 (1987) 253.
%
\bibitem{quant} M.K. Gaillard, T.R. Taylor, Nucl. Phys. B 381 (1992)
577; V. S. Kaplunovsky, J. Louis, Nucl. Phys. B 444 (1995) 191;
M. K. Gaillard, B. Nelson, Y.Y. Wu, Phys. Lett. B 459 (1999) 549.
%
\bibitem{thresh} L.J. Dixon, V.S. Kaplunovsky and J. Louis, Nucl. Phys
B 355 (1991) 649; I. Antoniadis, K.S. Narain and T.R. Taylor,
Phys. Lett.  B 267 (1991) 37.
%
\bibitem{yy} Y.-Y. Wu, hep-th/9610089.
%
\bibitem{agnt} I. Antoniadis, E. Gava, K.S. Narain and T.R. Taylor,
Nucl. Phys. B 476 (1996) 133.
%
\bibitem{mk} M. K. Gaillard, hep-th/0412079.
%
\bibitem{joel} J. Giedt, Ann. of Phys. (N.Y.) 297 (2002).
%
\bibitem{bdfs} F. Buccella, J.-P. Derendinger,
S. Ferrara and C. A. Savoy, {\it Phys. Lett.} {B 115} 375 (1982);
M.~K. Gaillard and J. Giedt, {\it Phys. Lett.} {\bf B479}
  308 (2000).
%
\bibitem{gn} J. Giedt and B. D. Nelson, JHEP 0405 (2004) 069.
\end{thebibliography}
\end{document}